\pdfoutput=1
\documentclass[aps,prl,reprint,superscriptaddress,showpacs,amsmath]{revtex4-1}
\usepackage[T1]{fontenc}
\usepackage[utf8]{inputenc}

\usepackage{isotope}
\usepackage{xcolor}
\usepackage{graphicx}
\usepackage{tikz}
\usepackage{hyperref}

\usepackage{times,txfonts}
\usepackage{url}

\newcommand{\LOsixdash}[1][]{\tikz[baseline=-2.5pt]{\draw[line width=1.5pt,#1] (0,0) -- +(8pt,0pt);}}

\newcommand{\LOsevendash}[1][]{\tikz[baseline=-2.5pt]{\draw[line width=1.5pt, dash pattern=on 2.67pt off 2.67pt,#1] (0pt,0pt) -- (8pt,0pt);}}

\newcommand{\circlemark}[1][]{\tikz[baseline=-2.5pt]{\draw[#1] (0,0) circle[radius=2pt];}}
\newcommand{\squaremark}[1][]{\tikz{\draw[#1] (0,0) rectangle (4pt,4pt);}}

\newcommand{\alphaNN}{\ensuremath{\alpha_{N}}}
\newcommand{\alphaYN}{\ensuremath{\alpha_{Y}}}

\begin{document}

\title{\emph{Ab Initio} Description of p-Shell Hypernuclei}

\author{Roland Wirth}
\email{roland.wirth@physik.tu-darmstadt.de}
\affiliation{Institut f\"ur Kernphysik, Technische Universit\"at Darmstadt, Schlossgartenstr. 2, 64289 Darmstadt, Germany}

\author{Daniel Gazda}
\affiliation{ECT*, Villa Tambosi, 38123 Villazzano (Trento), Italy}
\affiliation{Nuclear Physics Institute, 25068 \v{R}e\v{z}, Czech Republic}

\author{Petr Navr\'atil}
\affiliation{TRIUMF, 4004 Wesbrook Mall, Vancouver, British Columbia, V6T 2A3, Canada}

\author{Angelo Calci}
\author{Joachim Langhammer}
\author{Robert Roth}
\email{robert.roth@physik.tu-darmstadt.de}
\affiliation{Institut f\"ur Kernphysik, Technische Universit\"at Darmstadt, Schlossgartenstr. 2, 64289 Darmstadt, Germany}

\date{\today}

\begin{abstract}
We present the first \emph{ab initio} calculations for p-shell single-$\Lambda$ hypernuclei. For the solution of the many-baryon problem, we develop two variants of the no-core shell model with explicit $\Lambda$ and $\Sigma^+,\Sigma^0,\Sigma^-$ hyperons including $\Lambda$-$\Sigma$ conversion, optionally supplemented by a similarity renormalization group transformation to accelerate model-space convergence. In addition to state-of-the-art chiral two- and three-nucleon interactions, we use leading-order chiral hyperon-nucleon interactions and a recent meson-exchange hyperon-nucleon interaction. We validate the approach for s-shell hypernuclei and apply it to p-shell hypernuclei, in particular to $\isotope[7][\Lambda]{Li}$, $\isotope[9][\Lambda]{Be}$ and $\isotope[13][\Lambda]{C}$. We show that the chiral hyperon-nucleon interactions provide ground-state and excitation energies that generally agree with experiment within the cutoff dependence. At the same time we demonstrate that hypernuclear spectroscopy provides tight constraints on the hyperon-nucleon interactions.
\end{abstract}


\maketitle


Over the past decades, the structure of hypernuclei has been the focus of a number of experimental programs worldwide, providing a wealth of high-precision data on excitation spectra as well as binding energies \cite{Davis2005,Hashimoto2006,Gal2008,Gal2012,GiIm10,JuliaMa13}. These experimental efforts continue and are intensified, e.g., in several present and future experiments at international facilities like J-PARC, JLab, and FAIR.
Hypernuclear structure theory has a rich history of phenomenological models that have accompanied and driven the experiments, most notably, the shell model for p- and sd-shell hypernuclei \cite{Gal1971,*Gal1972,*Gal1978,Millener2008,*Millener2010,*Millener2012}, cluster models \cite{Motoba1983,Motoba1985,Hiyama2009,Hiyama2012}, various mean-field models \cite{GuDh12,Vidana2001,VrPo98,GlVo93}, or recent Monte Carlo calculations with simplified phenomenological interactions \cite{LoGa13,LoPe14}. \emph{Ab initio} calculations based on realistic nucleonic and hyperonic interactions were limited to systems of up to 4 nucleons so far \cite{Nemura2002,Nogga2002,Haidenbauer2007,Nogga2013}. Nevertheless, these calculations established a direct link between experimental observables and the underlying interactions and helped to elucidate the role of hyperons in matter. Advancing \emph{ab initio} methods beyond their current limits is highly desirable. It would allow to exploit the wealth of accurate experimental data, e.g., on p-shell hypernuclei, for constraining and improving the underlying interactions and to make predictions for yet unobserved phenomena.

There are two main aspects that hindered \emph{ab initio} calculations for p-shell hypernuclei in the past. Firstly, a prerequisite are accurate \emph{ab initio} calculations of the non-strange parent nucleus. The approach has to be able to provide converged results for the parent nucleus and the nucleonic Hamiltonian has to yield a good description of the experimental nuclear spectra. In the past few years, \emph{ab initio} methods using two-nucleon (NN) and three-nucleon (3N) interactions constructed in chiral effective field theory (EFT) succeeded to provide a quantitative description of ground states and spectra of nuclei in the p-shell and beyond \cite{RoLa11,RoCa14}. This is facilitated by a multitude of developments on computational many-body methods that give access to an unprecedented range of nuclei \cite{RoBi12,BiLa14,HeBi13,TsBo11,SoBa13,CiBa13}.

Secondly, the hyperon-nucleon (YN) interaction is ill constrained due to the scarce scattering data in the YN sector. Different models for the YN interaction, such as the widely used NSC and ESC models of the Nijmegen group \cite{RiNa10,RiYa06}, quark models \cite{Fujiwara1996a,*Fujiwara1996b} or the J\"ulich meson exchange models \cite{Haidenbauer2005}, yield different results at the level of cross sections already, rendering a meaningful \emph{ab initio} description of hypernuclei difficult. In a new development, chiral EFT has been employed to derive YN interactions within the same conceptual framework as the nucleonic interactions. Leading-order (LO) and, very recently, next-to-leading order (NLO) chiral YN interactions were developed by Polinder \emph{et al.} \cite{Polinder2006} and Haidenbauer \emph{et al.} \cite{Haidenbauer2013b}, respectively, succeeding their earlier meson-exchange interactions like the J\"ulich'04 model \cite{Haidenbauer2005}. An exciting option for constraining YN interactions directly from QCD emerges from recent lattice QCD calculations \cite{BeCh12,Sasa13}, e.g., for YN phase shifts. In combination with the advances in \emph{ab initio} many-body methods, this opens unique opportunities to learn about the structure of hypernuclei from first principles. By confronting accurate calculations with precise hypernuclear data, one can characterize and constrain the YN interaction, which is still the main source of uncertainty, and assess the relevance of three-baryon interactions for hypernuclear structure. Quantitative knowledge of the two- and three-baryon interactions is vital to understand not only hypernuclear structure but also the role of hyperons in dense baryonic matter in connection with the structure of neutron stars \cite{BaHa12,BuSc11,ViLo11}.

In this Letter, we present the first \emph{ab initio} calculations for p-shell single-$\Lambda$ hypernuclei.
We employ two versions of the no-core shell model (NCSM) for the solution of the many-body problem \cite{NaVa00,BaNa13}, the Jacobi NCSM (J-NCSM) and the importance truncated NCSM (IT-NCSM). We include nucleons, the $\Lambda$, and all $\Sigma$ hyperons as explicit degrees of freedom, thus accounting for the full $\Lambda$-$\Sigma$ coupled-channel problem. In both approaches we employ the same NN and 3N interactions derived in chiral EFT. We use the chiral NN interaction at N$^3$LO by Entem \& Machleidt \cite{Entem2003} and the local form of the chiral 3N interaction at N$^2$LO \cite{Navratil2007} with low-energy constants determined from $A=3$ binding energies and triton half-life \cite{GaQu09}, both for $500\,\text{MeV}/c$ cutoff momentum. In the YN sector we employ the J\"ulich'04 interaction \cite{Haidenbauer2005} as a representative for the meson-exchange models and the LO chiral YN interaction \cite{Polinder2006} with cutoff momenta of $600$ and $700\,\text{MeV}/c$ to probe the cutoff dependence. The hypernuclear Hamiltonian is transformed via a similarity renormalization group (SRG) evolution to accelerate the convergence of the NCSM-type many-body calculations.

\emph{Many-body method.}
The NCSM provides an extremely versatile framework for the formulation of an \emph{ab initio} method for hypernuclei. We have developed two independent but equivalent variants: (\emph{i}) the J-NCSM using a harmonic-oscillator (HO) basis in relative Jacobi coordinates \cite{NaKa00}, which enables an explicit center-of-mass separation and allows for calculations up to large numbers of HO excitation quanta, defining the basis-truncation parameter $N_{\max}$, for three- and four-baryon systems. (\emph{ii}) the IT-NCSM using a basis of Slater determinants of HO single-particle states with an optional importance truncation of the $N_{\max}$ model space \cite{Roth09,RoCa14}, which allows us to treat hypernuclei throughout the whole p-shell and beyond. Both approaches include nucleons and the $\Lambda$ and $\Sigma^+,\Sigma^0,\Sigma^-$ hyperons explicitly with their physical rest masses \cite{Nakamura2010}. The many-baryon model spaces are constrained by the total baryon number $A$, the electric charge $Q$, and the strangeness $\mathcal{S}$, thus, the full coupled-channel problem including $\Lambda$-$\Sigma$ conversion and explicit $\Sigma$ baryons is solved. Furthermore, all Coulomb interactions as well as the charge symmetry breaking terms of the NN and YN interaction are included.

\emph{Similarity renormalization group.}
In order to accelerate the convergence of the NCSM calculations with model-space size, we optionally employ an SRG transformation of the Hamiltonian \cite{BoFu10,RoNe10,RoCa14,JuNa09}, which has been very successful in the context of \emph{ab initio} nuclear structure calculations \cite{RoLa11,RoBi12,RoCa14}. This specific unitary transformation is based on the flow equation $\mathrm{d}H_{\alpha}/\mathrm{d}\alpha = [\eta_\alpha, H_{\alpha}]$ using the dynamic generator $\eta_\alpha = m_N^2 [T_{\text{int}}, H_\alpha]$, with the intrinsic kinetic energy $T_\text{int}$, the evolved Hamiltonian $H_\alpha$, and the flow parameter $\alpha$. The flow equation is solved numerically in a momentum or HO basis. We use an explicit particle representation, again accounting for all possible channel couplings resulting from tensor-type interactions, the antisymmetric spin-orbit terms, and the $\Lambda$-$\Sigma$ conversion as well as for the different rest masses. Furthermore, we introduce different flow parameters $\alphaNN$ and $\alphaYN$ for channels involving only nucleons and channels involving a hyperon, respectively. For purely nucleonic channels we perform the evolution in two- and three-particle space, giving access to the SRG-evolved NN and 3N interactions, which is state of the art for nuclear structure calculations \cite{RoCa14}.

For channels involving hyperons, we are presently limited to evolutions in two-body space, thus YNN interactions formally induced by the SRG transformation cannot be included directly. However, a variation of the flow parameters $\alphaNN$ and $\alphaYN$ probes the effect of induced YNN interactions---this is completely analogous to the use of the flow parameter as a diagnostic tool for induced 3N and 4N interactions in nucleonic systems \cite{RoLa11,RoCa14}. We find that the SRG evolution of YN channels generates large induced YNN interactions, whereas the evolution in NN channels only yields a weak flow-parameter dependence. Therefore, we restrict ourselves to $\alphaYN=0\,\text{fm}^4$ in the following. We will discuss the origin of the strong induced YNN interactions as well as their physical impact in a separate publication.

\begin{figure}
  \includegraphics[width=0.95\columnwidth]{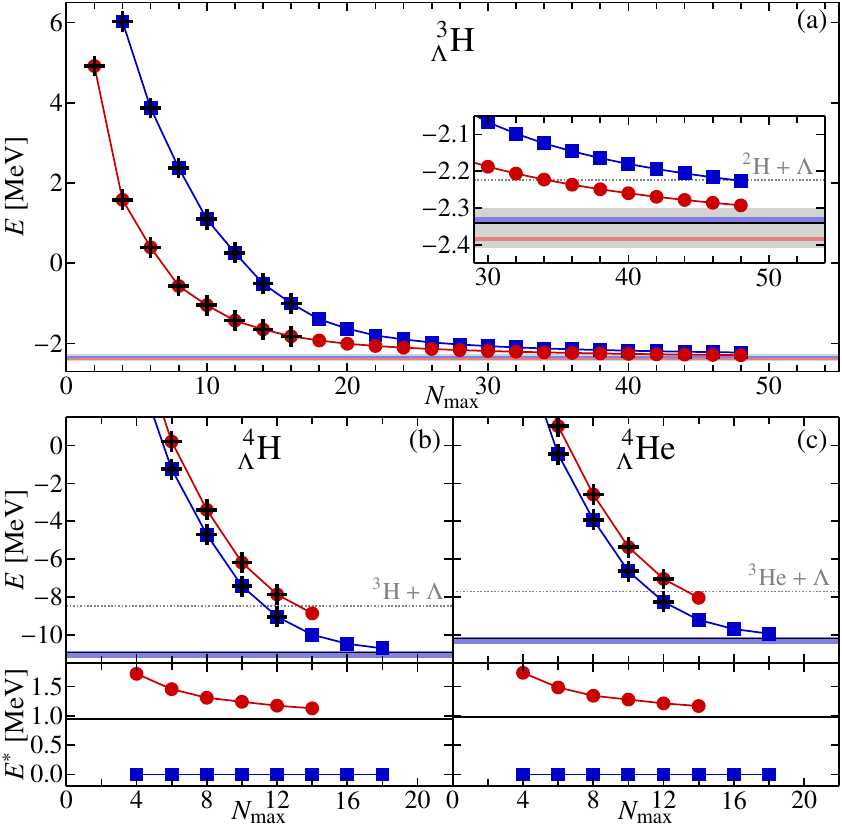}
    \vskip-2ex
  \caption{\label{fig:lh3}(color online) Ground-state energy of s-shell hypernuclei obtained with the LO chiral YN interaction with cutoff $600\,\text{MeV}/c$. Solid symbols represent J-NCSM results, crosses show IT-NCSM results. Panel (a) shows the ground-state energies of $\isotope[3][\Lambda]{H}$ for $\hbar\Omega=20\,\text{MeV}$, $\alphaYN=0\,\text{fm}^4$ and $\alphaNN=0\,\text{fm}^4$(\squaremark[fill]) and $\alphaNN=0.08\,\text{fm}^4$ (\circlemark[fill]) with EFT-motivated extrapolations (colored bands) compared to the experimental value (gray band) and the result of a Faddeev calculation \cite{Haidenbauer2007} (\LOsixdash, see inset). Panels (b) and (c) show results for the $0^+$ ground states (\squaremark[fill]) and $1^+$ first excited states (\circlemark[fill]) of $\isotope[4][\Lambda]{H}$ and $\isotope[4][\Lambda]{He}$, respectively, using $\alphaYN=\alphaNN=0\,\text{fm}^4$ and $\hbar\Omega=28\,\text{MeV}$. The upper plots show absolute energies, the lower plots excitation energies. The colored bands give the result of an exponential extrapolation of the ground-state energy and the solid lines represent results of previous few-body calculations \cite{Haidenbauer2007}.}
\end{figure}

\emph{Validation for s-shell hypernuclei.}
In a first step we validate the two NCSM implementations for the s-shell hypernuclei \isotope[3][\Lambda]{H} and \isotope[4][\Lambda]{H}, \isotope[4][\Lambda]{He}, where exact few-body calculations using the same YN interactions are available. Figure \ref{fig:lh3} shows the $N_{\max}$-dependence of the ground-state energies obtained in the J-NCSM and the IT-NCSM for the LO chiral YN interaction with cutoff $600\,\text{MeV}/c$ compared to results from Faddeev calculations \cite{Haidenbauer2007}. For \isotope[3][\Lambda]{H} we observe an extremely slow convergence related to the weak binding. However, the large $N_{\max}$ spaces accessible with the J-NCSM in combination with recent EFT-motivated extrapolation schemes for weakly bound states (see Eq. (44) of Ref. \cite{More2013}, using between 5 and 10 data points for the largest $N_{\max}$ to extract nominal value and uncertainty) yield a ground-state energy of $-2.33(1)\,\text{MeV}$ using the bare Hamiltonian in excellent agreement with the result of Ref. \cite{Haidenbauer2007}. There is a tiny difference of the extrapolated energies for $\alphaNN=0$ and $0.08\,\text{fm}^4$ of about $50$ keV, hinting at a small effect of induced YNN terms resulting from the SRG evolution of the nucleonic channels. For \isotope[4][\Lambda]{He} and \isotope[4][\Lambda]{H} the $N_{\max}$ convergence is much better, even with the bare Hamiltonian including chiral NN, 3N and YN interactions, as shown in Fig. \ref{fig:lh3}(b). The energies for the $0^+$ ground state obtained from an exponential extrapolation using between 3 and 6 data points for the largest $N_{\max}$ are $-11.1(1)$ and $-10.3(1)$ MeV for \isotope[4][\Lambda]{H} and \isotope[4][\Lambda]{He}, respectively, corresponding to $\Lambda$ separation energies of $2.6(1)$ MeV for both nuclei, consistent with Ref. \cite{Haidenbauer2007}. The excitation energies of the $1^+$ excited states, as shown in the lower plots, also agree very well with previous few-body calculations and with experiment. Both NCSM approaches agree at the level of $1-5\,\text{keV}$ in all model spaces accessible to both, thus validating the implementations.

\begin{figure}
  \includegraphics[width=0.9\columnwidth]{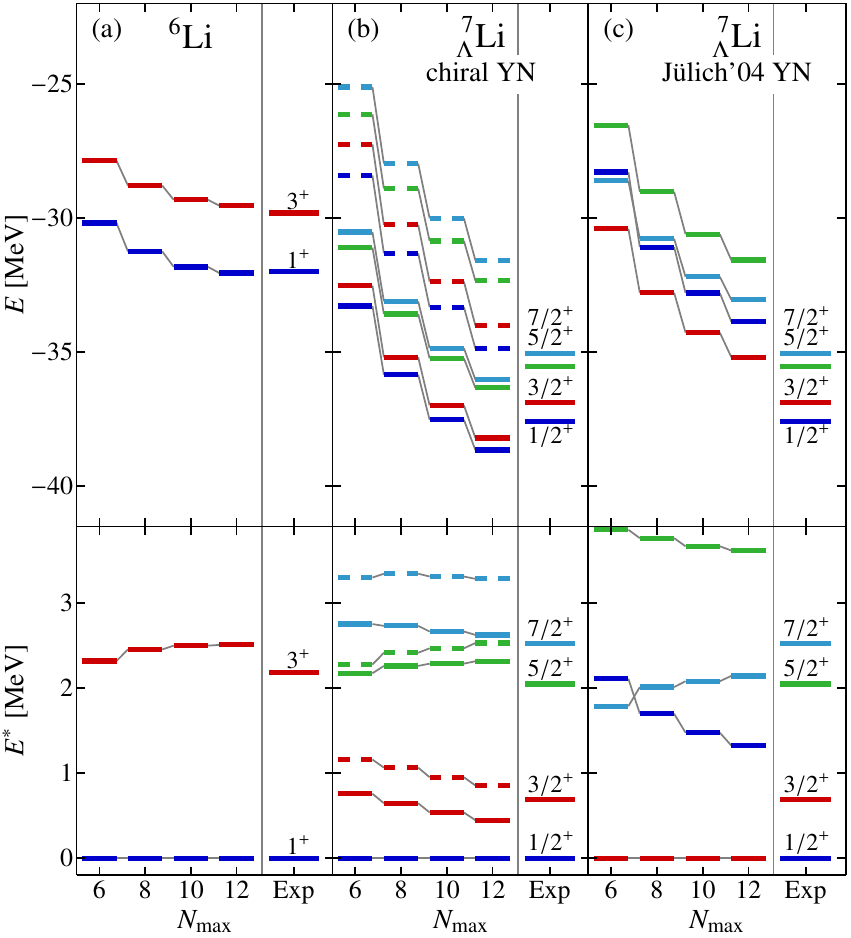}
    \vskip-2ex
  \caption{\label{fig:lli7}(color online) Absolute and excitation energies of the first four states of \isotope[7][\Lambda]{Li} for the LO chiral (b) and the J\"ulich'04 YN interaction (c) compared to the non-strange parent nucleus \isotope[6]{Li} (a). For the LO chiral YN interaction in panel (b) we use the two cutoff values $600\,\text{MeV}/c$ (\LOsixdash) and $700\,\text{MeV}/c$ (\LOsevendash).
    Experimental data from Refs.~\cite{Davis2005,Hashimoto2006,NuDat2}. All calculations use $\alphaNN=0.08\,\text{fm}^4$, $\alphaYN=0.0\,\text{fm}^4$, and $\hbar\Omega=20\,\text{MeV}$.
  }
\end{figure}

\emph{Application to p-shell hypernuclei.}
The IT-NCSM enables \emph{ab initio} calculations for all single-$\Lambda$ hypernuclei throughout the p-shell. Here we focus on a representative subset, where precise experimental data on the spectroscopy is available. We discuss \isotope[7][\Lambda]{Li} as one of the best studied p-shell hypernuclei in both experiment and phenomenological models, \isotope[9][\Lambda]{Be} for which the first spin-doublet is degenerate posing a fine-tuning problem for the interaction, and \isotope[13][\Lambda]{C} representing the upper p-shell. In comparison to the well studied s-shell, hypernuclei in the p-shell probe higher relative partial waves of the YN interaction and thus enhance spin-orbit and tensor effects. Based on these calculations we assess the performance of present YN interactions, in particular, the J\"ulich'04 and the LO chiral YN interactions for cutoff momenta $600$ and $700\,\text{MeV}/c$.

We start with the discussion of \isotope[7][\Lambda]{Li} in Fig. \ref{fig:lli7}. Panel (a) shows the absolute energies and the excitation energies of the non-strange parent nucleus \isotope[6]{Li} obtained for the chiral NN+3N interaction with an SRG evolution to $\alphaNN=0.08\,\text{fm}^4$. Note that the converged energies are practically independent of $\alphaNN$ in the lower p-shell \cite{RoCa14,RoLa11}. The good agreement of absolute and excitation energies with experiment resulting from the chiral NN+3N Hamiltonian and the good convergence of the IT-NCSM are evident and a prerequisite for accurate hypernuclear calculations.

When adding a hyperon to the non-strange parent nucleus, in a simple picture, the weak attractive YN interaction leads to a lowering of the ground-state energy and to a splitting of each $J>0$ level into a doublet with angular momenta $J+\frac{1}{2}$ and $J-\frac{1}{2}$. The energy splitting is directly controlled by and sensitive to the YN interaction. Both effects are evident in the IT-NCSM results for \isotope[7][\Lambda]{Li} in panels (b) and (c) of Fig. \ref{fig:lli7}. Moreover, the differences between the YN interactions are evident. For the J\"ulich'04 interaction employed in Fig. \ref{fig:lli7}(c) the ground-state energy is in reasonable agreement with experiment, but the level ordering is wrong. The splitting of the spin-doublet is significantly too large and has the wrong sign, leading to a systematically reversed level ordering. This deficiency is already visible for the excited states of the $A=4$ hypernuclei \cite{Haidenbauer2007}.

The LO chiral YN interactions employed in Fig. \ref{fig:lli7}(b) provide a consistently better description of the spectra. The ground-state energies obtained for cutoff 600 and 700 MeV/$c$ are slightly below and above experiment, respectively. The excitation energies exhibit a weaker cutoff dependence, with the cutoff 600 MeV/$c$ yielding slightly lower excitation energies. If we interpret this dependence on the YN cutoff as an estimator for the effects of higher-order terms in the chiral expansion, then we can state that the LO chiral YN interaction gives ground-state and excitation energies that agree with experiment within the truncation uncertainties.

\begin{figure}
  \includegraphics[width=0.9\columnwidth]{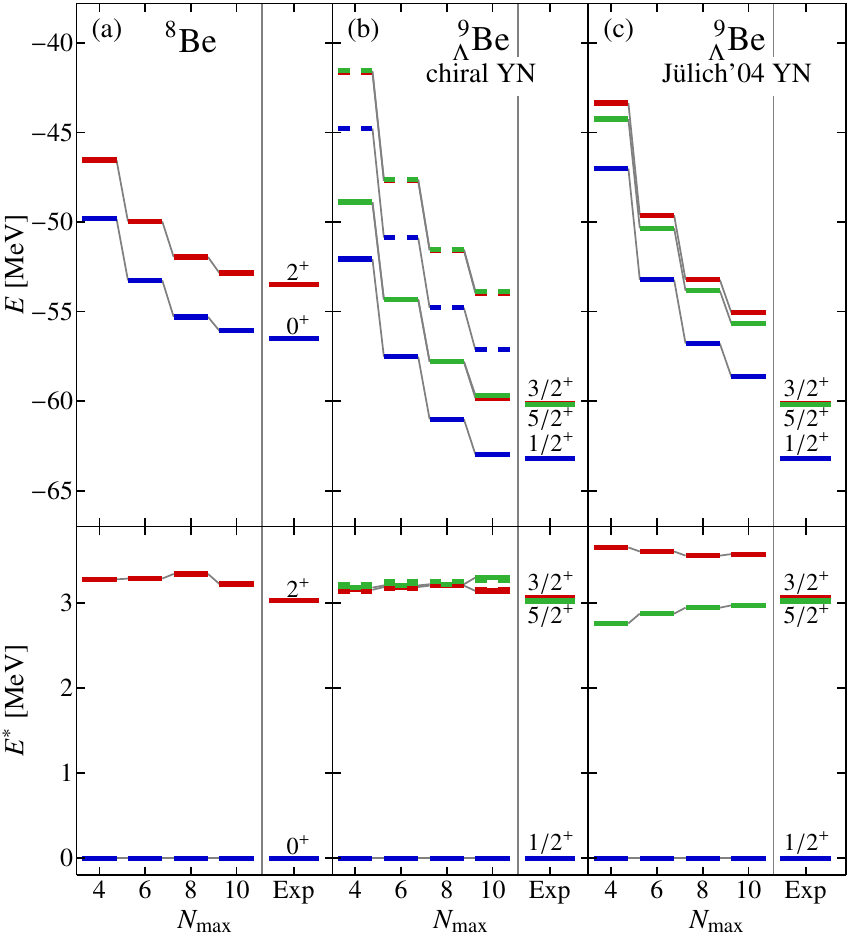}
    \vskip-2ex
  \caption{\label{fig:lbe9}(color online) Same as Fig. \ref{fig:lli7}, but for \isotope[9][\Lambda]{Be} and \isotope[8]{Be}.}
\end{figure}

The IT-NCSM also gives access to spectroscopic observables such as transition strengths. As an example we consider the $B(\text{E2})$ strength for the $5/2^+\to1/2^+$ transition in \isotope[7][\Lambda]{Li}, which has been experimentally determined to $B(\text{E2})=3.6^{+0.5}_{-0.5}\text{(stat)}^{+0.5}_{-0.4}\text{(syst)}\,e^2\text{fm}^4$ \cite{Tanida2001}. For the LO chiral YN interation with cutoff 600 MeV/$c$ we obtain $B(\text{E2})=2.3(1)$ and $2.4(1)\,e^2\text{fm}^4$ for $N_{\max}=10$ and $12$, respectively, using $\hbar\Omega=20\,\text{MeV}$. The numbers in brackets indicate the uncertainties of the threshold extrapolation \cite{RoCa14}. Obviously, convergence of this long-range observable is problematic and a systematic study exploiting the frequency-dependence to perform extrapolations is needed. A simpler example is the $B(\text{M1})$ strength for the spin-flip transition $3/2^+\to1/2^+$. We obtain $B(\text{M1})=0.31(1)\,\mu_N^2$ for $N_{\max}=10$ and $12$, indicating good convergence. This is in excellent agreement with a preliminary experimental value reported in \cite{Tamura2006}.

As a second case we discuss the spectrum of \isotope[9][\Lambda]{Be} as depicted in Fig. \ref{fig:lbe9}. The nucleonic parent nucleus \isotope[8]{Be} is unbound with respect to decay into two $\alpha$-particles, but still the IT-NCSM provides a good description of the ground- and excited-state energies in a bound-state approximation. The addition of the hyperon binds the \isotope[9][\Lambda]{Be} hypernucleus. Again the LO chiral YN interactions for cutoff 600 and 700 MeV/$c$ yield different ground-state energies that bracket the experimental value. A peculiarity of \isotope[9][\Lambda]{Be} is that the spin-doublet resulting from the $2^+$ state in \isotope[8]{Be} is practically degenerate, with the higher-$J$ state being at slightly lower excitation energy experimentally, contrary to the other light hypernuclei. The LO chiral YN interactions reproduce the excitation energy of the doublet and the near degeneracy within threshold extrapolation and convergence uncertainties. In contrast, the J\"ulich'04 interaction gives a significant splitting of the spin doublet in contradiction to experiment.

\begin{figure}
  \includegraphics[width=0.9\columnwidth]{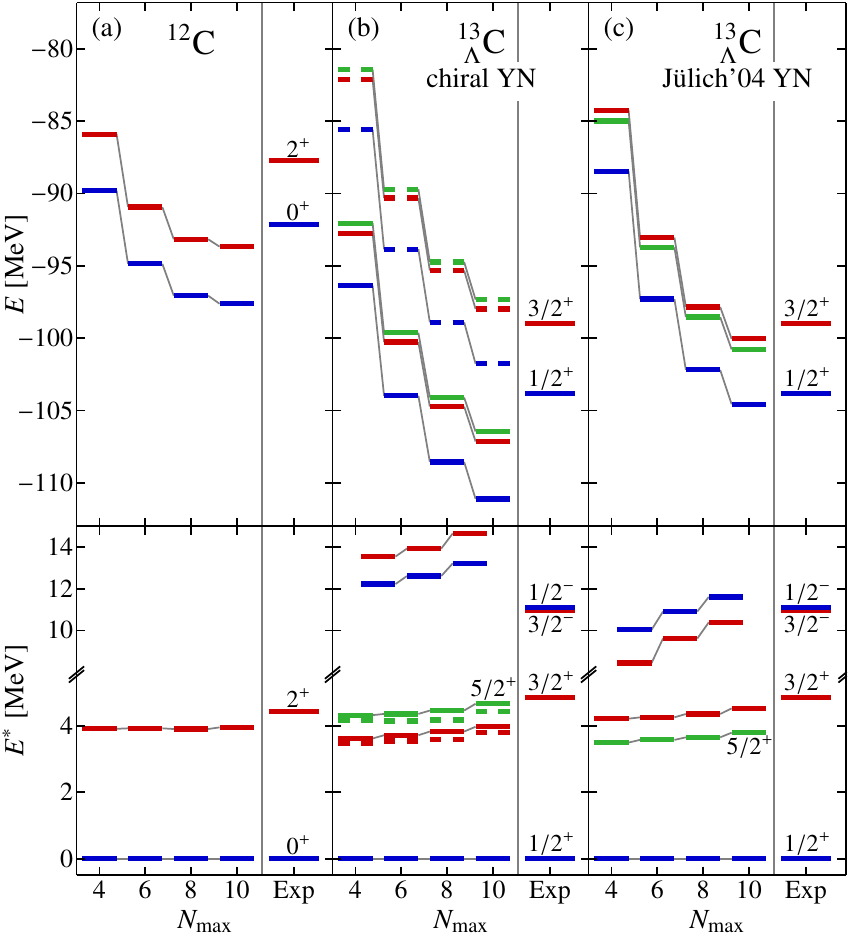}
    \vskip-2ex
  \caption{\label{fig:lc13}(color online) Same as Fig. \ref{fig:lli7}, but for \isotope[13][\Lambda]{C} and \isotope[12]{C}. Note the change of scale in the lower panels.}
\end{figure}

As a final example from the upper p-shell we discuss \isotope[13][\Lambda]{C} in Fig. \ref{fig:lc13}. The SRG-evolved chiral NN+3N interaction at $\alphaNN=0.08\,\text{fm}^4$ gives a ground-state energy of the nucleonic parent \isotope[12]{C} about $6\,\text{MeV}$ below experiment. This overbinding is related to the emergence of SRG-induced 4N interactions in the upper p-shell that are not included in the present calculations (see Refs. \cite{RoCa14,RoLa11}). The absolute energies of \isotope[13][\Lambda]{C} inherit this overbinding, however, taking this into account, the chiral LO interactions are consistent with the experimental ground-state energies within the cutoff uncertainty. Also the excited spin-doublet appears at a slightly too low excitation energy, since the $2^+$ excited state in \isotope[12]{C} is already too low. The splitting of the spin doublet is predicted by the LO chiral YN interactions to be $650$ to $700$ keV for the largest model spaces. Again, the J\"ulich'04 YN interaction predicts the opposite level ordering for the doublet. Note that the $5/2^+$ state was not yet observed experimentally, but cluster-model calculations \cite{Hiyama2009} put it below the $3/2^+$ state in contrast to the LO chiral YN interaction. We also calculated the lowest doublet of unnatural parity states, shown in the lower panels of Fig. \ref{fig:lc13}, which are dominated by a hyperon in a p-orbit. Neither the chiral nor the J\"ulich'04 YN interaction can reproduce the near degeneracy of the $1/2^-$ and $3/2^-$ states as observed experimentally. This hints at deficiencies in higher partial waves, which are strongly affected by sub-leading contributions to the chiral YN interactions.

\emph{Conclusions.}
We have performed the first \emph{ab initio} calculations for single-$\Lambda$ p-shell hypernuclei using NCSM approaches with explicit hyperons. After a validation for s-shell hypernuclei, we have studied selected p-shell hypernuclei using J\"ulich'04 and the LO chiral YN interactions. Within the expected cutoff dependence the LO chiral YN interactions reproduce the experimental data up to the mid-p-shell, whereas the J\"ulich'04 YN interaction systematically gives wrong orderings and splittings of the spin-doublet states. For \isotope[13][\Lambda]{C} the situation is unclear as the $5/2^+$ state is not known experimentally. Neither of the YN interactions describes the first negative-parity doublet correctly, which hints at deficiencies in the higher relative partial-waves. This illustrates the potential of systematic \emph{ab initio} studies of p-shell hypernuclei for improving our understanding of the YN interaction. In this context, the inclusion and validation of the chiral YN interactions at NLO is highly desirable. At the same time the impact of SRG-induced and initial chiral YNN interactions needs to be investigated.

\begin{acknowledgments}

We thank A. Nogga and J. Haidenbauer for useful discussions and for providing us with the YN interaction codes.
This work is supported by DFG through SFB 634, by the Helmholtz International Center for FAIR (HIC for FAIR), by the BMBF (06DA7047I), by the GACR Grant No. 203/12/2126, by the EU initiative FP7, HadronPhysics3, under the SPHERE and LEANNIS cooperation programs, and by the NSERC Grant No. 401945-2011. TRIUMF receives funding via a contribution through the Canadian National Research Council.
The authors gratefully acknowledge computing time granted by the J\"ulich Supercomputing Center (JUROPA), the CSC Frankfurt (LOEWE-CSC), and the computing center of the TU Darmstadt (LICHTENBERG).

\end{acknowledgments}

%

\end{document}